%% file: af_lett.tex
\documentclass[aps,prb,preprint]{revtex4-1}%
\usepackage{amsmath}
\usepackage{graphicx}
\usepackage{amssymb}

\usepackage{color} 
\usepackage{graphicx}
\usepackage{bm}
\usepackage{amsmath}

\usepackage{gt17}

\DeclareMathAlphabet\mathbfcal{OMS}{cmsy}{b}{n}
\DeclareMathAlphabet{\mathpzc}{OT1}{pzc}{m}{it}
\begin{document}
\title{
Theory of spin transport through antiferromagnetic insulator
}
\author{Gen Tatara} 
\author{Christian Ortiz Pauyac}
\affiliation{RIKEN Center for Emergent Matter Science (CEMS)
and RIKEN Cluster for Pioneering Research (CPR), 
2-1 Hirosawa, Wako, Saitama, 351-0198 Japan}

\date{\today}

\begin{abstract}
A theoretical formulation for spin transport through an antiferromagnetic (AF) insulator is presented in the case driven/detected by direct/inverse spin Hall effect in two heavy metal contacts. 
The spin signal is shown to be transferred by the ferromagnetic correlation function of the antiferromagnet, which is calculated based on a magnon representation. To cover high temperature regimes, we include an auxiliary field representing short AF correlations and a temperature-dependent damping due to magnon scattering.  
The diffusion length for spin is long close to the degeneracy of the two AF magnons, and has a maximum as function of temperature near the N\'eel transition. 
\end{abstract}  

\maketitle
\newcommand{\lambdaso}{\lambda_{\rm so}}
\newcommand{\elle}{\ell_{\rm e}}
\newcommand{\spinx}{1}
\newcommand{\spiny}{2}
\newcommand{\spinz}{3}
\newcommand{\CAF}{C}
\newcommand{\kmax}{k_{\rm max}}
\newcommand{\omegamax}{\omega_{\rm max}}
\newcommand{\al}{a_{0}}
Spin current injection to various materials has been a hot issue in spintronics.
Of particular recent interest is spin current propagation in antiferromagnetic insulators (AFI).
Being common material, AFI have practical advantages in materials choice.
Moreover, insertion of an antiferromagnetic (AF) layer between ferromagnet and normal metal was found to enhance spin current injection efficiency \cite{WangAF14,Lin16}. 
Experimentally, spin current injection and propagation efficiency in AF insulators is reported to vanish or very small at $T=0$, to have a peak near the N\'eel transition temperature, $T_{\rm N}$, reducing at higher temperatures \cite{Lin16,Cramer18}.

Transmission of spin information in antiferromagnets  is an intriguing issue as fundamental science. 
For describing spin current injection to antiferromagnets, two issues need to be clarified, namely, to what degree of freedom the incident spin current couples, and how it propagates.
Obviously, rigid AF order does not react to spin current injection having a particular spin polarization, and fluctuation is essential.
There are two branches of AF magnons, corresponding to opposite spin angular momentum, and coupling of the two modes is essential as noted previously \cite{Rezende16,Khymyn16}. 
The amplitude and decay length of spin current propagation are expected to depend strongly on the temperature because of the Bose distribution function representing the number of AF magnon excitations. 
In fact, spin current amplitude was found experimentally to have peak near $T_{\rm N}$ and this feature was explained based on a phenomenological theory using mixing conductance \cite{Lin16}. 
A sharp peak at $T_{\rm N}$ was predicted in another theory evaluating fluctuations around mean-field solution in the spatially uniform case \cite{Okamoto16}.
The frequency-dependence of  magnon propagation length was theoretically studied in Ref. \cite{Cramer18}, although the relation between AF magnon propagation and spin current propagation remained untouched. 

The objective of the present paper is to provide a transparent formalism to describe propagation of spin information through an AF insulator. 
We do not rely on the conventional spin current picture, as it is ambiguous due to non-conservation of spin current.
Moreover, introducing phenomenological parameters such as spin mixing conductance makes straightforward understanding of phenomena difficult. 
Here we follow the linear response theory for the applied electric field treating the exchange interaction between spins in heavy metals and antiferromagnets perturbatively. 
The description is an application of Ref. \cite{TataraISH18} indicating that spin current propagation is equivalent to correlation function of ferromagnetic (FM) spin fluctuation or magnetic susceptibility. 
Ferromagnetic fluctuation of antiferromagnets is represented by an exchange or pair 
creation/annihilation of two AF magnons.
The spin information is therefore transferred by magnon pair correlation propagator, just in the same manner as magnetic susceptibility in FM metals is represented by an electron-hole pair propagation. 
The mismatch of frequencies of FM excitation of GHz and of AF one of THz therefore does not matter as the magnon pair correlation can absolve or emit low external frequencies.
Moreover, shortening of AF  correlation at high temperature does not necessarily block spin current propagation, because FM fluctuation on the contrary grows.
Instead, magnon lifetime at high temperatures is greatly reduced by strong magnon scattering \cite{Harris71}, resulting in a significant reduction of spin current propagation. 
As a result, the propagation efficiency has a peak near $T_{\rm N}$, although the peak position, determined by the competition of the fluctuation and damping, does not necessarily coincides with $T_{\rm N}$.

The correlation of spin transport with magnetic correlation has been pointed out experimentally in Refs. \cite{WangAF15,Lin16,Qiu16}. 
In the case of spin pumping into a heavy metal, the efficiency of spin current injection was argued to be determined by the imaginary part of magnetic susceptibility of heavy metal divided by external angular frequency \cite{Ohnuma14}, although  their treatment of external angular frequency was theoretically not comprehensive

\begin{figure}[bt]
 \includegraphics[width=0.4\hsize] {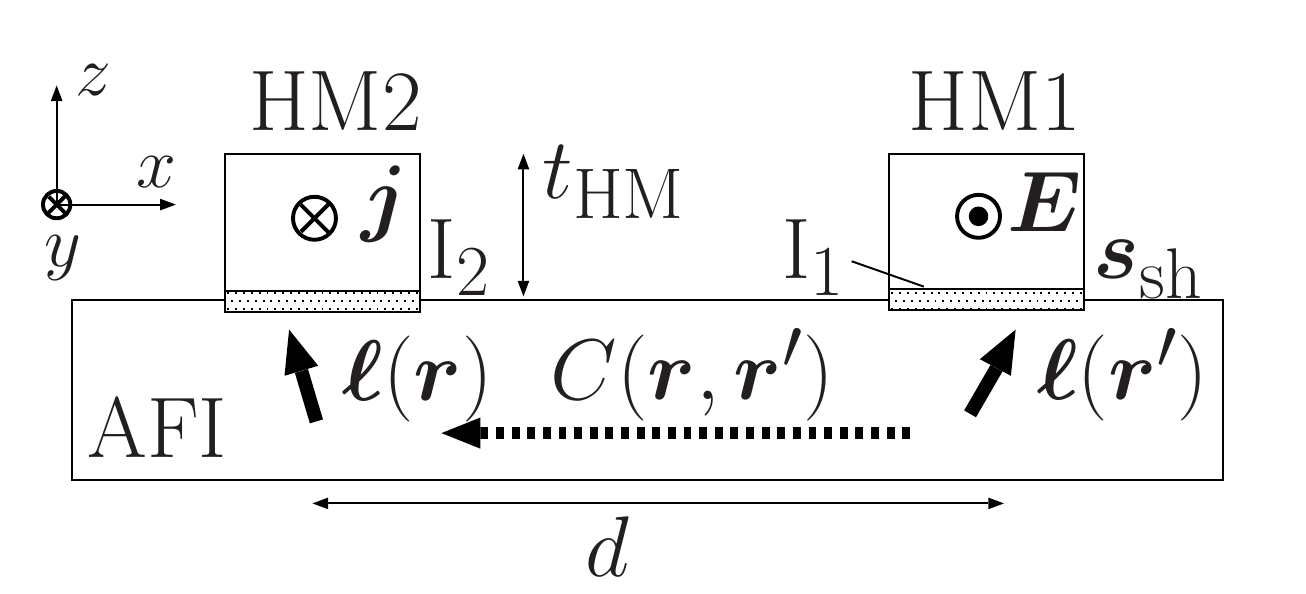}
\caption{Setup of nonlocal direct and inverse spin Hall effects. HM1 and HM2 are heavy metal leads for spin current  generation and detection, respectively. In the conventional picture, 'spin current' generated by spin Hall effect in HM1 is transmitted through AFI and measured at HM2 by the inverse spin Hall effect. 
The coupling between HM1 and HM2 at interfaces ${\rm I}_1$ and ${\rm I}_2$, respectively, is induced by interface exchange interaction $J_{\rm I}$ between spin polarization in HMs and FM spin component $\ellv$ of AFI. 
'Spin current propagation' through AFI is expressed by the FM correlation function $C(\rv,\rv')$ of $\ellv$.
The moment $\ellv$ is related to the N\'eel vector $\nv$ as $\ellv\propto \nv\times\dot{\nv}$, and it turns out that spin information is transferred only if $\nv\parallel\hat{\xv}$ in the present geometry.
\label{FIGsetup}}  
\end{figure}

Let us develop  step by step a linear response theory to describe the nonlocal direct and inverse spin Hall effects separated by an AFI (Fig. \ref{FIGsetup}). (See Supplementary material for details.)
The key interaction is the coupling between heavy metals (HMs) and AFI at interfaces.
Here we consider an $sd$ exchange interaction between electron spin polarization in HM and FM component of AFI, namely
\begin{align}
 H_{{\rm I}} &= \sum_{i=1,2} J_{\rm I} \int_{I_i}\frac{d^3r}{a^3} \ellv \cdot (c^\dagger \sigmav c), \label{HeAF}
\end{align}
where  $I_{i}$ ($i=1,2$) denotes the interface between HM$i$ and AFI, $J_{\rm I} $ is a coupling strength, $\sigmav$ denotes the Pauli matrix and $c^\dagger$ and $c$ are electron field operators. 
The FM moment of AFI is defined as $\ellv\equiv (\Sv_{\rm A}+\Sv_{\rm B})/S$, where $\Sv_{\rm A}$ and $\Sv_{\rm B}$ are spin on the two sublattices A and B, respectively 
($S\equiv|\Sv_{\rm A}|=|\Sv_{\rm B}|$). 
This coupling is natural as the first approximation as the electron wave function overlap would smear out the staggered (N\'eel) component of localized spin in AFI.
Let us start with the inverse spin Hall effect (ISHE) in HM2. 
In the context of linear response theory, the driving field of ISHE is a non-equilibrium FM moment $\ellv$ of AFI in Eq.  (\ref{HeAF}).
The output electric current is thus described by a correlation function of current and spin density, 
$\tilde{\chi}^{JS}_{ik}$, as $j_i=J_{\rm I}\tilde{\chi}^{JS}_{ik}\ell_k$ (suppressing spatial coordinates).
(Exact expression is presented in the Supplementary material.)
The FM moment $\ellv$ near I$_{2}$ is generated non-locally by the spin Hall effect (SHE) in HM1. 
The  SHE is described by the correlation function $\tilde{\chi}^{SJ}$, the reciprocal of $\tilde{\chi}^{JS}$, as 
$s_{{\rm sh},l}= \tilde{\chi}^{SJ}_{lj}E_j$, where $s_{{\rm sh},l}$ and $E_j$ are the spin density induced by SHE and the applied electric field in HM1, respectively \cite{TataraISH18}.
Taking account of $H_{\rm I}$, the FM moment $\ellv$ induced near I$_2$  by the spin accumulation at I$_1$ as a result of SHE is written using nonlocal FM spin correlation function $\CAF(\rv)$ as 
$\ell_k(\rv_1)=J_{\rm I}\int_{{\rm I}_2}d^3r_2 \CAF_{kl}(\rv_1-\rv_2) \tilde{\chi}^{SJ}_{lj}(\rv_2)E_j(\rv_2)$, where subscripts $k$ and $l$ denote spin direction.
As FM moment is expressed as a composite field of two AF magnons, the correlation function $\CAF(\rv)$ is a two magnon propagator as we shall see below.

Summarizing, the inverse spin Hall current is represented as a product of three correlation functions as 
\begin{align}
 j_i (\rv) &= (J_{I})^2 \int_{I_1}d^3r_1 \int_{I_2}d^3r_2 \int_{\rm HM1} d^3r'\tilde{\chi}^{JS}_{ij}(\rv-\rv_2) \CAF_{kl}(\rv_2-\rv_1) \tilde{\chi}^{SJ}_{lm}(\rv_1-\rv') E_m(\rv').
 \label{allcorrelation}
\end{align}
The correlation functions in Eq. (\ref{allcorrelation}) turn out to be the physical correlation function ${\chi}^{JS}$ determined by the lesser component divided by the external frequency $\Omega$, i.e., 
$\tilde{\chi}^{JS}_{ij}\equiv -\lim_{\Omega\ra0} \frac{1}{i\Omega}{{\chi}^{JS}_{ij}(\Omega)}$. (See Supplementary material.)
The correlation function ${\chi}^{JS}$  is linear in $\Omega$ because equilibrium spin accumulation does not generate electric current that is dissipative, and thus $\tilde{\chi}^{JS}$ has a static component.
Moreover, considering HM as a bulk, inversion symmetry is present and the spatially-uniform component of the current-spin correlation vanishes, meaning that $\chi^{JS}_{ij}$ 
starts from the first order in the external wave vector $\qv$ \cite{TataraISH18}.   
Thus, direct and inverse spin Hall effects with current perpendicular to the spin accumulation profile is described in the ballistic case  by the correlation function 
$\tilde{\chi}^{JS}_{ij}(\qv,\Omega)=i\lambda_{\rm sh} \epsilon_{ijk} q_k$, where  $\epsilon_{ijk}$ is the totally antisymmetric tensor.
A coefficient $\lambda_{\rm sh}$, determined by the spin-orbit interaction strength is related to  dimensionless spin Hall angle $\theta_{\rm sh}(\equiv \js/j)$  as 
$\theta_{\rm sh}=\lambda_{\rm sh}/(\sigmab\taue)$, where $\sigmab$ and $\taue$ are the Boltzmann conductivity and elastic electron lifetime, respectively \cite{TataraISH18}. 
Taking account of diffusive electron motion in HMs, the function is multiplied by a diffusion factor 
$D_{\rm s}(\qv)\equiv \frac{1}{Dq^2\tau+\gamma_{\rm sf}}$, where  $D$ is a diffusion constant, $\gamma_{\rm sf}$ is related to a static spin diffusion length $\ell_{\rm sf}$ in HM as $\ell_{\rm sf}=\sqrt{3} \elle /\sqrt{\gamma_{\rm sf}}$,  $\elle=\kf\taue/m$ being electron elastic mean free path, as \cite{TataraISH18} 
\begin{align}
 \tilde{\chi}_{ij}^{JS}(\qv) 
& =   \lambda_{\rm sh} \epsilon_{ijk} i q_k D_{\rm s}(q).
\label{chijsresy}
\end{align}
The current is therefore expressed as 
\begin{align}
 j_i (\rv)
 &= (\lambda_{\rm sh}J_{I})^2 \epsilon_{ijk}\epsilon_{lmn} \int_{I_1}d^3r_1\int_{I_2} d^3r_2\nabla^{\rv}_j D_{\rm s}(\rv-\rv_2) \int_{\rm HM1}d^3r' \CAF_{kl}(\rv_2-\rv_1)\nabla^{\rv'}_m D_{\rm s}(\rv_1-\rv') E_n(\rv') ,  \label{jresultdif}
\end{align}
where the spin diffusion propagator is 
$D_{\rm s}(\rv)=\frac{3\ell_{\rm sf}\al}{2\elle^2}e^{-\frac{r}{\ell_{\rm sf}}}$ ($\al$ is the lattice constant of HM). 
Spatial derivative of spin diffusions in Eq. (\ref{jresultdif}) represents spin current flow of the conventional picture, as spin current is proportional to a gradient of spin density in the diffusive regime.
In the common setup in Fig. \ref{FIGsetup}, the derivatives are in the perpendicular direction, which we choose as the $z$ direction.
The derivative at I$_1$ of HM1 is evaluated as $\nabla_z D_{\rm s}|_{r=0}= -\frac{3\al}{2\elle^2}$.
For HM2, we discuss the averaged current for the thickness of HM2, $t_{\rm HM}$, i.e., 
$\overline{j}\equiv\frac{1}{t_{\rm HM}}\int_0^{t_{\rm HM}}dz j(z)$, where we use  
$\frac{1}{t_{\rm HM}}\int_0^{t_{\rm HM}}dz \nabla_z D_{\rm s}(z)
=-\frac{3\ell_{\rm sf}\al}{2t_{\rm HM}\elle^2}(1-e^{-{t_{\rm HM}}/{\ell_{\rm sf}}})$.

The correlation function of AFI, $\CAF_{kl}$ is calculated later and we proceed here using the results. 
It turns out to vanish for spin direction perpendicular to the N\'eel vector, $\nv$, and the spatial dependence is exponential in the most cases.  We denote the direction in the spin space of AFI by $(\spinx, \spiny,\spinz)$ to remember that spin space is independently of coordinate space, and $\nv$ is chosen along $\spinz$-direction. As shown below, the correlation function of AFI is 
$\CAF_{kl}= \delta_{k\spinz}\delta_{l\spinz}C(\rv)$, where 
 \begin{align}
    C(\rv) &=c_0\frac{a }{2\xi} e^{-|\rv|/\xi}, \label{Cr}
 \end{align}
with a dimensionless constant $c_0$ and a FM correlation length $\xi$ ($a$ is the lattice constant of AFI).
Chosing the applied current direction as the $y$ axis (Fig. \ref{FIGsetup}), the antisymmetric tensors in Eq. (\ref{jresultdif}) indicate that   'spin current' propagates only if $\nv(=\hat{\bm{\spinz}})=\hat{\bm{x}}$.

\begin{figure}
 \includegraphics[width=0.45\hsize] {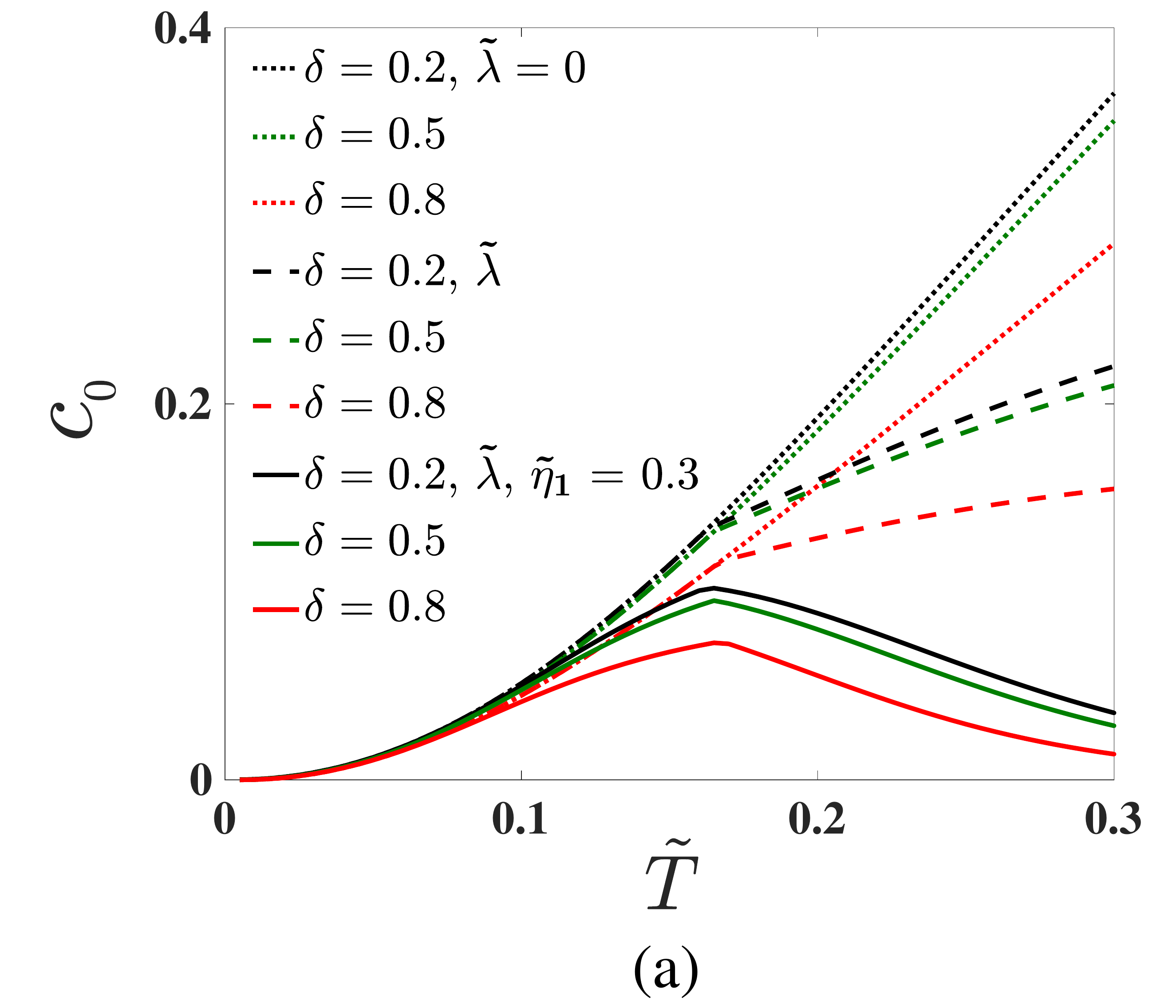}
 \includegraphics[width=0.45\hsize] {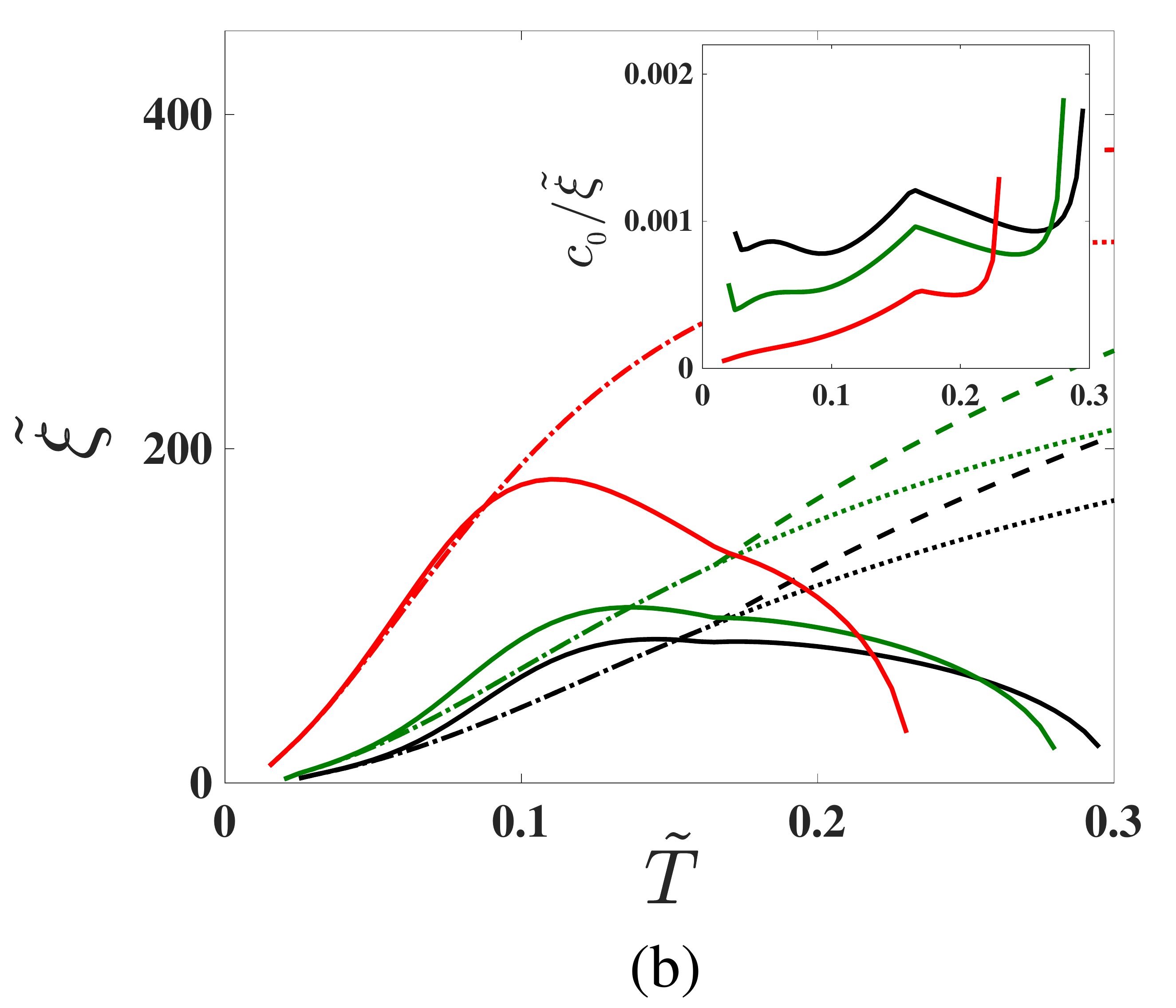}
\caption{ Plots of (a) $c_0$ and (b) dimensionless FM correlation length $\tilde{\xi}\equiv \xi/a$  as functions of normalized temperature, $\tilde{T}\equiv \kb T/\omegamax$, where $\omegamax\equiv v\kmax$ is the maxmim magnon energy, $\kmax\equiv \pi/a$. 
Dotted lines are without auxiliary field $\lambda(\equiv\tilde{\lambda}\omegamax^2)$ and for $\eta_1=0$, dashed lines are with $\lambda$ and for $\eta_1=0$, and solid lines are physical ones including both $\lambda$ and $\tilde{\eta_1}\equiv \eta_1/\omegamax=0.3$. 
Bare damping is $\eta_0/\omegamax=10^{-4}$, and the two energy gaps are  $\Delta$ and $\Delta \delta$, with $\Delta/\omegamax=0.03$ and 
$\delta=0.2,0.5$ and $0.8$. The N\'eel temperature in the present model is $\tilde{T}_{\rm N}\simeq0.16$. 
The inset of (b) shows the ratio $\frac{c_0}{\tilde{\xi}}$, which governs the amplitude of ISH signal (Eq. (\ref{javratio})). Anomalous behaviors in the high-temperature regime with $\xi\lesssim 0$ indicate breakdown of our model. 
\label{FIGc0xi}}
 \end{figure}

The ISH current is opposite to the applied electric field.
Defining  an effective nonlocal conductivity  $\overline{\sigma}$ as 
$ \overline{j}\equiv -\overline{\sigma}E$, we have
 \begin{align}
 \frac{\overline{\sigma}}{\sigmab} 
 &= \frac{3}{8}(\theta_{\rm sh})^2 (J_{I})^2 \dos \tau  \lt( \frac{\al}{\elle} \rt)^2 
 \frac{c_0 a}{\xi} e^{-d/\xi}\frac{\ell_{\rm s}}{t_{\rm HM}}(1-e^{-t_{\rm HM}/\ell_{\rm s}})
 ,  \label{javratio}
\end{align}
where  $d$ is the distance between HM1 and HM2, and 
$\dos$ is the electron density of states. 
The electron properties are insensitive to the temperature around room or lower temperatures.
According to analysis below, the ratio $\frac{c_0 a}{\xi}$ of AFI does not depend much on the temperature either, as both $c_0$ and $\xi$ have similar temperature profiles (Fig. \ref{FIGc0xi}(b)), and thus the dominant temperature dependence is expected to arise from $ e^{-d/\xi(T)}$. 
Equation (\ref{javratio}) indicates that the interface exchange coupling constant $J_{\rm I}$ can be determined experimentaly from the magnitude of ISHE. 

\begin{figure}
 \includegraphics[width=0.4\hsize] {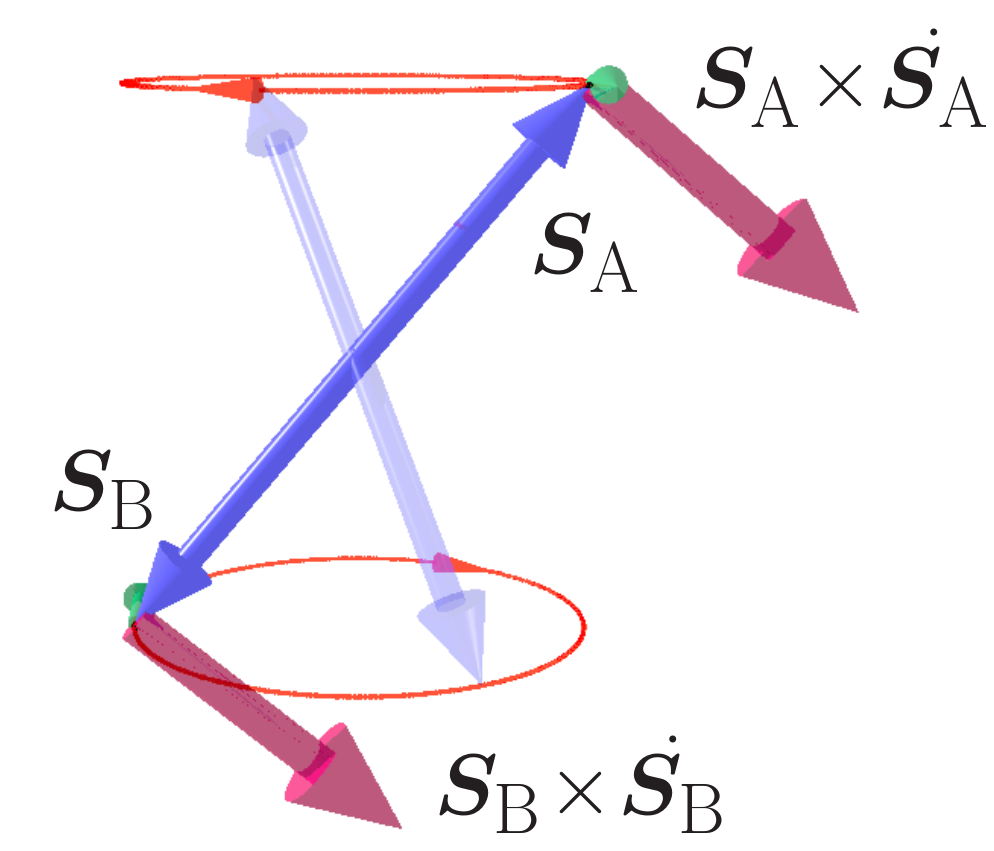}
\caption{Schematic picture showing that the FM moment $\ell$ is proportional to $\nv\times\dot{\nv}$.  For both sublattices A and B  with opposite spin, $\Sv_{\rm A}\simeq S\nv$ and $\Sv_{\rm B}\simeq -S\nv$, $\Sv_{\rm A}\times\dot{\Sv_{\rm A}}$ and $\Sv_{\rm B}\times\dot{\Sv_{\rm B}}$ (largte arrows) point the same direction.
\label{FIGAF_precession}}  
\end{figure}
Let us start study of the correlation in AFI. 
We first note that the FM moment is expressed by $\nv$ as 
$\ellv =\frac{1}{6SJ_0}(\nv\times\dot{\nv})$, where $J_0$ is the AF exchange coupling. 
This relation, rigorously derived in the Supplementary materials, is understood by noting that $\Sv\times\dot{\Sv}$ points the same direction for the spins of both sublattices A and B (Fig. \ref{FIGAF_precession}).
The N\'eel vector has a classical expectation value $n_3$ along the direction $\hat{\bm \spinz}$  below the AF transition temperature $T_{\rm N}$. 
Fluctuation is represented by a two-component AF magnon field $\varphiv$ as 
$\nv=(\varphi^{(\spinx)},\varphi^{(\spiny)}, n_3)$, neglecting the second order of magnon field.
The FM moment around $\nv$ is therefore represented by a combination of the two magnons as 
$\ell_\spinz=\varphi^{(\spinx)} \dot{\varphi}^{(\spiny)}-\dot{\varphi}^{(\spinx)} {\varphi}^{(\spiny)}$, while the orthogonal components are linear in magnon field.
Considering the fact that AF dynamics (typically in the THz regime) is much faster than the FM one (GHz), only the moment $\ell_\spinz$ has a low energy coupling to AF magnons.
Namely, the spin polarization parallel to $\nv$ can be transported for a long distance, while AFI does not react to the perpendicular polarization, resulting in  
$\CAF_{kl}(\qv) \equiv \delta_{k\spinz}\delta_{l\spinz} C_\qv$. 
This feature  is in agreement with recent experiment \cite{Lebrun18}.
The two magnon modes $\varphi^{(\spinx)}$ and $\varphi^{(\spiny)}$ carries the opposite angular momentum, and thus FM moment is induced by an exchange of the two modes (ordinary process)  or by a pair annihilation or creation (anomalous processes). 
In terms of magnon creation/annihilation operators $a^{(i)}$ and $a^{(i)\dagger}$ ($i=1,2$), introduced as 
$ \varphi_i(k)=\sqrt{\frac{g}{ \omega_k^{(i)} }}(a^{(i)}_k+a^{(i)\dagger}_{-k})$ ($g=3J_0$),  the expectation value of induced moment with wave vector $\qv$ reads 
\begin{align}
 \average{ \ell_\spinz(\qv,t)} 
  &= \frac{1}{2S} \sum_{\kv} \frac{1}{\sqrt{\omega^{(1)}_{\kv}\omega^{(2)}_{\kv+\qv}}} 
  \nnr
 & \times 
 \lt[ (\omega^{(2)}_{\kv+\qv}-\omega^{(1)}_{\kv}) 
  [F_{-\kv,\kv+\qv}(t,t) - \overline{F}_{\kv,-(\kv+\qv)}(t,t)  ]
  +(\omega^{(2)}_{\kv+\qv}+\omega^{(1)}_{\kv}) 
  [   D^{(21)}_{\kv+\qv,\kv}(t,t) - D^{(12)}_{-\kv,-(\kv+\qv)}(t,t)  ] 
  \rt],
\end{align}
where $\omega^{(i)}_k\equiv\sqrt{(vk)^2+(\Delta^{(i)})^2}$ is the magnon energy for branch $i$ ($v$ and $\Delta^{(i)}$ being the magnon velocity and gap, respectively) and  
\begin{align}
  F_{-\kv,\kv+\qv}(t,t') &\equiv -i \average{ a^{(1)}_{-\kv}(t)a^{(2)}_{\kv+\qv}(t')} , &
  D^{(ij)}_{\kv+\qv,\kv}(t,t') &\equiv -i \average{a^{(i)}_{\kv+\qv}(t) a^{(j)\dagger}_{\kv}(t')}   ,\label{DFdefs}
\end{align}
 are anomalous and ordinary path-ordered Green's functions on a complex time path  and $  \overline{F}\equiv F^*$.
The static component of $\average{ \ell_\spinz}$ induced by SHE in HM1 is written using a correlation function $\CAF_\qv$ as $\average{ \ell_\spinz(\qv)}\equiv J_{\rm I} \CAF_{\qv} s_{{\rm sh},\spinz}(\qv)$, where 
\begin{align}
\CAF_{\qv}&=
   \frac{g}{2S} \sum_{\kv} \Re\biggl[  \frac{1}{\omega^{(1)}_{\kv}\omega^{(2)}_{\kv+\qv}}
  \nnr
 & \times \lt(
  (1+n_k^{(1)}+n_{\kv+\qv}^{(2)}) 
  \frac{
 (\omega^{(2)}_{\kv+\qv}-\omega^{(1)}_{\kv})^2}{\omega_{\kv+\qv}^{(2)}+\omega_{\kv}^{(1)}-i (\eta_k +\eta_{k+q})}
-(n_{\kv+\qv}^{(2)}-n_k^{(1)}) 
 \frac{(\omega^{(2)}_{\kv+\qv}+\omega^{(1)}_{\kv})^2 }{\omega_{\kv+\qv}^{(2)}-\omega_{\kv}^{(1)}-i (\eta_k +\eta_{k+q})}
  \rt) \biggr]  \label{Cqgeneral}.
\end{align}
Here $n_k^{(i)}\equiv [{e^{\beta \omega_k^{(i)}}-1}]^{-1}$ is Bose distribution function ($\beta\equiv1/(\kb T)$, $\kb$ being the Boltzmann constant), $\eta_k$ represents magnon damping and $\Re$ denotes the real part.
The first term of the right-hand-side of Eq. (\ref{Cqgeneral}) without Bose distribution function is the quantum contribution that exists at $T=0$. 
Spin current can thus transmit though antiferromagnet at $T=0$, where no magnons are excited.
(The quantum pair creation process has been shown to be essential for the neutron scattering of Haldane antiferromagnets at $T=0$ \cite{Affleck92}.) 

The correlation function determines the spatial profile of steady 'spin current propagation'. 
Long-range behavior is determined by the small $q$ behavior,
\begin{align}
\CAF_{q}
& = c_0+ c_2 q^2+O(q^4).
\end{align}
For the degenerate case, $\omega_k^{(1)}=\omega_k^{(2)}$, the uniform contribution $c_0$ vanishes.
When the two spin waves have different gaps as a result of magnetic anisotropy (like in the case of NiO), the uniform component $c_0$ is finite, which leads to efficient 'spin current propagation'.
The length scale of the spin information propagation, a diffusion length of spin, is given by  
$\xi\equiv \sqrt{-\frac{c_2}{c_0}}$, as the response function is approximately written as 
$\CAF_{q} \simeq  \frac{c_0}{1+\xi^2 q^2}+O(q^4)$, which leads in the real space to an exponential decay within a distance of $\xi$, Eq. (\ref{Cr}).

Fig. \ref{FIGc0xi} shows numerical results of $c_0$ and $\xi$ as function of temperature in nondegenerate cases with the two energy gaps $\Delta^{(1)}=\Delta$ and $\Delta^{(2)}=\Delta \delta$. 
Close to the degeneracy, $\delta\sim1$, spin transport is long-ranged (larger $\xi$) as the transport is mediated by the mixing of the two magnon branches. 
In contrast, $c_0$ representing the magnitude of spin transmission is suppressed for larger $\delta$, simply due to an increase of $\Delta^{(2)}$. 

Magnon representation is usually applied to low temperatures compared to $T_{\rm N}$.
However, the representation itself does not necessarily break down even above $T_{\rm N}$ as far as short-ranged AF correlation persists for a length larger than the lattice constant, just like the case of FM magnons well-defined in the presence of structures like a domain wall.  
Short-ranged correlation is theoretically described by introducing an auxiliary field $\lambda(T)$ to impose the constraint $|\nv|=1$ by the saddle point approximation\cite{Chakravarty89}.
The field contributes to a temperature-dependent gap and modifies 
the magnon dispersion to be  $\omega_k^{(i)}= \sqrt{v^2k^2+(\Delta^{(i)})^2+\lambda}$.
The static AF correlation length,  
$\xi_{\rm AF}^{(i)}=v/\sqrt{(\Delta^{(i)})^2+\lambda+\eta_k^2}$ including damping $\eta_k$, is usually shorter than the FM correlation length governing spin propagation. (See the  Supplementary Material.)
The auxiliary field description is known to describe well the AF correlation length above $T_{\rm N}$ \cite{Chakravarty89,Yamamoto91}.

What is most essential for transport at high temperatures is the magnon damping due to magnon interactions at high density.
The effect of magnon interaction on the damping was studied theoretically in detail in Ref. \cite{Harris71}.
It was shown that the scattering induces a self energy proportional to $T^3$ and $\omega_k^2$ for low-energy magnon. 
We here include the effect in the damping constant $\eta$ as 
\begin{align}
 \eta(T,k)=\eta_0+ \eta_1 \lt(\frac{k}{\kmax}\rt)^2 \lt(\frac{\kb T}{\omegamax}\rt)^3 ,
\end{align}
where $\eta_0$ and $\eta_1$ are  constants, $\kmax\equiv\pi/a$ and $\omegamax\equiv v\kmax$ are the cutoffs for high wavevector and energy, respectively. 
The spin propagation efficiency $c_0$ and FM correlation length $\xi$ are significantly suppressed by the temperature-dependent damping at high temperature as seen in Fig. \ref{FIGc0xi} (solid and dashed lines).
The peak temperature, determined by the competition between magnon excitation number and damping, is close to $T_{\rm N}$.

Considering the case of NiO, $\omegamax/(2\pi)=30$ THz, $\Delta^{(1)}=1$ THz, $\Delta^{(2)}=0.2$ THz and $a=4.2$\AA \cite{Rezende16}, and our calculation ($\tilde{\Delta}=0.03$, $\delta=0.2$) indicates the spin transport length $\xi$ of the order of 20 nm around room temperature, which appears to be roughly consistent with experiment indicating the diffusion length of 10nm \cite{Rezende16}. 
For quantititave calculation, however, our model assuming square lattice with the nearest-neighbor hopping is too simple and more realistic modeling is necessary.

We have presented a theoretical formulation of spin injection into antiferromagnetic insulator (AFI) using direct and inverse spin Hall effects. 
The 'spin current propagation', induced by a AF magnon pair propagation, was shown to be represented by a ferromagnetic (FM) correlation function, $\CAF_{\qv}$, or a $q$-resolved FM susceptibility. 
Although $\CAF_\qv$ may appear similar to the conductivity for spin current in the analogy with the case of charge current, this is not the case because a correlation function of spin current representing the spin current conductivity can not be written by a spin correlation as spin is not conserved. 
The correlation function was studied based on magnon representation including an auxiliary field in the stationary-field approximation to cover the temperatures above the N\'eel temperature $T_{\rm N}$.
The decay length of spin propagation $\xi$ was calculated from a pair propagation process for the non-degenerate case.
It is different from (longer than) AF correlation length $\xi_{\rm AF}$ determined by individual AF magnon propagation, similarly to the electron case where transport lengths are longer than the elastic mean free path.
$\xi(T)$  has a peak near $T_{\rm N}$ as a result of suppression due to the damping arising from magnon scattering at high temperatures.
The dominant temperature dependence of the spin propagation efficiency through AFI for a distance of $d$ is thus expected to be $e^{-d/\xi(T)}$.

\acknowledgements
GT thanks T. Ono, E. Saitoh, R. Lebrun and M. Kl\"aui 
for valuable discussions. 
He is  grateful for the Graduate School Materials Science in Mainz (MAINZ) for financial support (DFG GSC 266).
This work was supported by 
a Grant-in-Aid for Exploratory Research (No.16K13853) 
and 
a Grant-in-Aid for Scientific Research (B) (No. 17H02929) from the Japan Society for the Promotion of Science 
and  
a Grant-in-Aid for Scientific Research on Innovative Areas (No.26103006) from The Ministry of Education, Culture, Sports, Science and Technology (MEXT), Japan.

\appendix

\input{af_e.bbl}
\end{document}

%% file: af_e.bbl
%